\documentclass{ws-mpla}
\usepackage[super]{cite}
\usepackage{graphicx}
\usepackage{hyperref}
\def\draftnote{}%
\usepackage[square, sort, numbers]{natbib}

\begin{document}

\markboth{Belyaev, Berezhnoy, Likhoded, Luchinsky}
{Comments on "Study of $J/\psi$ production in jets"}

\title{Comments on "Study of $J/\psi$ production in jets"}

\author{I. Belyaev}
\address{Institute of Theoretical and Experimental Physics (ITEP), Moscow, Russia\\Ivan.Belyaev@itep.ru}

\author{A.V. Berezhnoy}
\address{SINP of Moscow State University, 119991 Moscow, Russia\\Alexander.Berezhnoy@cern.ch}

\author{A.K. Likhoded}
\address{Institute for High Energy Physics NRC “Kurchatov Institute”, 142281, Protvino, Russia\\Anatolii.Likhoded@ihep.ru}

\author{A.V. Luchinsky}
\address{Institute for High Energy Physics NRC “Kurchatov Institute”, 142281, Protvino, Russia\\Alexey.Luchinsky@ihep.ru}

\maketitle


\begin{abstract}
Recent LHCb measurements of the $J/\psi$ meson production in jets is analyzed using fragmentation jet function formalism. It is shown that disagreement with theoretical predictions for distribution over the fraction of $J/\psi$ transverse momentum $z(J/\psi)$ in the cases of prompt production can be explained if one takes into account evolution of the fragmentation function and contributions from double parton scattering mechanism.

\end{abstract}


\section{Introduction}	

\begin{figure}
	\centering{\includegraphics[width=0.7\textwidth]{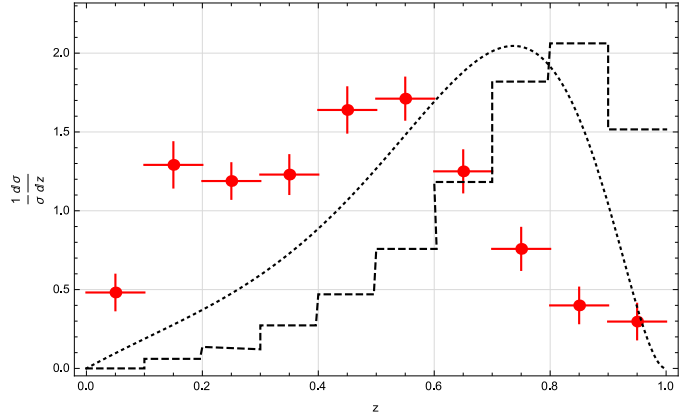}}
	\caption{$z(J/\psi)$ distribution of the promptly produced $J/\psi$ meson. Red points, dashed and dotted curves correspond to experimental data \cite{Aaij:2017fak}{}, Pythia8 predictions \cite{Aaij:2017fak}{}, and LO $c\to J/\psi$ fragmentation function \eqref{eq:FFB}{}. \label{Fig1}}
\end{figure}

In a recent experimental paper \cite{Aaij:2017fak} LHCb Collaboration analyzed $J/\psi$ meson production in induced by $c$-quark jets in the forward region of proton-proton interaction at center-of-mass energy $\sqrt{s}=13$ TeV. In the cited work distributions over fraction of the jet transverse momentum carried by the $J/\psi$ meson
\begin{align}
z &= \frac{p_T^{J/\psi}}{p_T^{jet}}
\end{align}
were presented both for $J/\psi$ mesons produced in $b$-hadron decays and promptly. It turns out, that in the first case experimental results are consistent with theoretical predictions, while in the latter case a noticeable disagreement with theory is observed. In Fig.\ref{Fig1} one can clearly see that measured by LHCb Collaboration $z$-distribution of promptly produced $J/\psi$ is significantly softer than theoretical predictions made by Pythia8 \cite{Sjostrand:2007gs} generator.
In our short note we will try to give a simple explanation of this disagreement.

\section{$J/\psi$ Production in Jets}

\begin{figure}
	\centering{\includegraphics[width=0.6\textwidth]{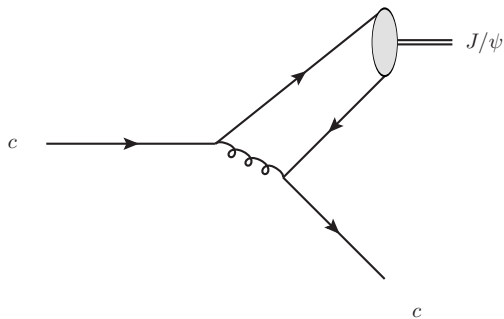}}
	\caption{Feynman diagram for $c\to J/\psi$ fragmentation\label{diag}}
\end{figure}

In the original experimental paper it is required that a transverse momentum of the jet $p_T^{jet}>p_T^\mathrm{min} = 20$ GeV, that is high enough to consider  the applicability of fragmentation approach. In this approach the measured distribution can be written in the form
\begin{align}
\frac{d\sigma_{J/\psi}}{dp_T} &=\int\limits^1_{2p_T/\sqrt{s}}
\frac{d\sigma_{c \bar c}}{dk_T}
\left(\frac{p_T}{z}\right)\frac{D_{c\to J/\psi}(z)}{z} dz,
\label{eq:frag_pT}
\\
\frac{1}{\sigma}\frac{d\sigma}{dz} &\sim D_{c\to J/\psi}(z),
\label{eq:FF}
\end{align}
where $D_{c\to J/\psi}(z)$ is the fragmentation function that describes $c$-quark transition into $J/\psi$ meson and  $k_T$ is a transverse momentum of $с$-quark. This function is universal and at LO QCD can be calculated using presented in Fig.\ref{diag} Feynman diagram.


  The analytical forms of fragmentation function for $S$ wave states are known from 
\cite{Braaten:1993jn,  Kiselev:1994qp}:

\begin{multline}
D_{Q \to (Q\bar q) }(z)=
\frac{2\alpha_s^2 |R_S(0)|^2}{27\pi m_c^3}
\frac{rz(1-z)^2}{(1-(1-r)z)^6}
(2-2(3-2r)z+3(3-2r+4r^2)z^2\\
-2(1-r)(4-r+2r^2)z^3+
(1-r)^2(3-2r+2r^2)z^4),
\label{eq:frag_vct}
\end{multline}
where  $\alpha_s$  is a strong coupling constant and $r=\frac{m_Q}{m_Q+m_q}$ and $|R_S(0)|$ is a value of $Q\bar{q}$ quarkonium wave function at origin.
 For our case $r=0.5$ and the formula can be rewritten as follows (see also \cite{Braaten:1993mp}):
 \begin{align}
D_{c\to J/\psi}(z) &\sim 
	\frac{4z(1-z)^2}{(2-z)^6}\left(16-32z+72z^2-32z^3+5z^4\right).
\label{eq:FFB}
\end{align}

Integrating over z, we can obtain the total  probability to produce $J/\psi$ and $c$ in fragmentation process~\cite{Braaten:1993mp}:
\begin{equation}
P_{c\to J/\psi}=\frac{64}{27 \pi}\alpha_s^2 
\frac{|R_S(0)|^2}{M_{J/\psi}^3} \Bigl( \frac{1189}{30}-57\log 2\Bigr).
\end{equation}
 
 As it is shown in \cite{Berezhnoy:1998aa}, the non-fragmentation mechanisms  essentially contribute to hadronic $J/\psi +c+\bar{c}$ production.  However these contributions rapidly decrease with transverse momenta increasing and  only mentioned above fragmentation contribution is left with close in rapidity $J/\psi$ meson and $c$ quark. This quark could be observed as $D$ meson in the same jet as $J/\psi$ and it is interesting to note that in our model we could expect $z(D)\approx z(J/\psi)/2$.
 
We show this function in Fig.\ref{Fig1} with dotted curve and one can see that its form (as well as Pythia8 predictions) contradicts experimental data.

\begin{figure}
	\centering{\includegraphics[width=0.7\textwidth]{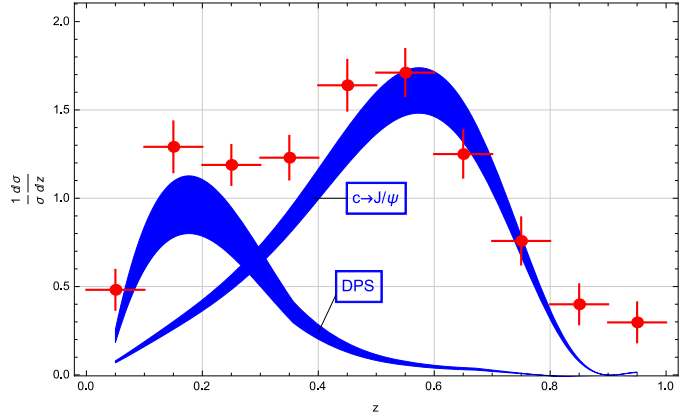}}
	\caption{
		Double parton scattering and fragmentation contributions to $J/\psi$ meson production in jets in comparison with experimental data. Theoretical uncertainties are from errors in \eqref{eq:c}.
		\label{Fig3}}
\end{figure}

It should be noted, however, that the fragmentation function \eqref{eq:FF} depends on the factorization scale $\mu^2$ and the parametrization \eqref{eq:FFB} corresponds to value $\mu=\mu_1=m_c$. Comparison with high-energy experimental results, on the other hand, should be performed at some higher scale of the order $\mu_2\sim p_T^\mathrm{min}$. From papers \cite{Novoselov:2010zz, Corcella:2007tg} it is known that the evolution from $\mu_1^2$ to $\mu_2^2$ leads to significant variation of the shape of fragmentation function. This evolution can be described using DGLAP evolution equation \cite{Gribov:1972ri, Lipatov:1974qm, Dokshitzer:1977sg, Altarelli:1977zs} and using it one can easy track the evolution of the fragmentation function \eqref{eq:FFB} from $\mu^2=\mu_1^2$ to $\mu^2=\mu_2^2$ scales. The corresponding results are shown in Fig.\ref{Fig3}. It is clear, that after the evolution of the fragmentation function was taken into account, the agreement with experimental data in high $z$ region is restored. There is, however, some experimental excess in $z\sim 0.2$ region. In order to describe this peak one should also add contributions from double parton scattering (DPS) mechanism. The corresponding distribution is presented in  \cite{Aaij:2017fak}{}. This distribution was obtained using Pythia8 generator with default setting, that in turn corresponds to $\sigma_\mathrm{eff}\sim 30$ mb. Simple estimates, on the other hand, shows, that experimental data supports higher contributions of the DPS mechanism (and, correspondingly, smaller value of the effective cross section), so in our work we use only the shape of DPS component.

\begin{figure}
	\centering{\includegraphics[width=0.7\textwidth]{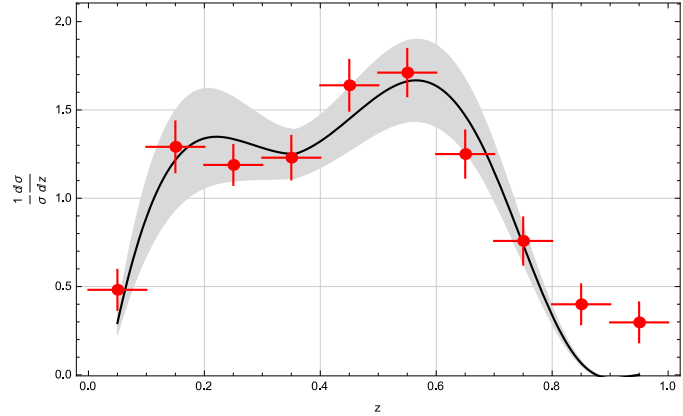}}
	\caption{
		Normalized $z$-distribution of total cross section in comparison with experimental data. Theoretical uncertainties are from errors in \eqref{eq:c}.
		\label{Fig4}}
\end{figure}

Since overall normalizations are not known, we describe total cross section as a sum of DPS and fragmentation signals with free parameters:
\begin{align}
\frac{1}{\sigma}\frac{d\sigma}{dz} &=
c_\mathrm{DPS}\left[\frac{1}{\sigma}\frac{d\sigma}{dz}\right]_\mathrm{DPS} +
c_\mathrm{frag}\left[\frac{1}{\sigma}\frac{d\sigma}{dz}\right]_\mathrm{frag},
\label{eq:total}
\end{align}
where numerical values of the parameters $c_{\mathrm{DPS},\mathrm{frag}}$ were determined from fit of the experimental data:
\begin{align}
c_\mathrm{frag} &= 0.59\pm0.05,\quad c_\mathrm{DPS} = 0.26 \pm 0.05.
\label{eq:c}
\end{align}
The correlation matrix of the fit is
\begin{align}
&\left(
\begin{array}{cc}
1 & -0.37 \\
-0.37 & 1
\end{array}
\right).
\end{align}
In Fig.\ref{Fig4} and Table \ref{tab:cs} we show calculated with these parameters total cross section in comparison with experimental data. One can clearly see that after double parton scattering contributions and evolution of the fragmentation function were taken into account, theoretical estimates are in reasonable agreement with the experiment.

\begin{table}[h]
\tbl{Normalized cross section}{
\begin{tabular}{*5l}
\toprule
$z$ & $c_\mathrm{frag}[\sigma^{-1}d\sigma/dz]_\mathrm{frag}$ & $c_\mathrm{DPS}[\sigma^{-1}d\sigma/dz]_\mathrm{DPS}$ & Total \eqref{eq:total} & Experiment \cite{Aaij:2017fak} \\
\hline
$0.05 \pm 0.05$ & $0.078 \pm 0.007$ & $0.08 \pm 0.02$ & $0.16 \pm 0.1$ & $0.48 \pm 0.1$  \\ 
$0.15 \pm 0.05$ & $0.28 \pm 0.03$ & $0.9 \pm 0.2$ & $1.2 \pm 0.3$ & $1.3 \pm 0.1$  \\ 
$0.25 \pm 0.05$ & $0.53 \pm 0.05$ & $0.78 \pm 0.1$ & $1.3 \pm 0.3$ & $1.2 \pm 0.1$  \\ 
$0.35 \pm 0.05$ & $0.87 \pm 0.08$ & $0.38 \pm 0.07$ & $1.3 \pm 0.2$ & $1.2 \pm 0.1$  \\ 
$0.45 \pm 0.05$ & $1.3 \pm 0.1$ & $0.15 \pm 0.03$ & $1.4 \pm 0.2$ & $1.6 \pm 0.1$  \\ 
$0.55 \pm 0.05$ & $1.6 \pm 0.1$ & $0.07 \pm 0.01$ & $1.6 \pm 0.3$ & $1.7 \pm 0.1$  \\ 
$0.65 \pm 0.05$ & $1.4 \pm 0.1$ & $0.034 \pm 0.007$ & $1.4 \pm 0.2$ & $1.3 \pm 0.1$  \\ 
$0.75 \pm 0.05$ & $0.73 \pm 0.07$ & $0.012 \pm 0.002$ & $0.74 \pm 0.1$ & $0.76 \pm 0.1$  \\ 
$0.85 \pm 0.05$ & $0.11 \pm 0.01$ & $0.0081 \pm 0.002$ & $0.12 \pm 0.02$ & $0.4 \pm 0.1$  \\ 
$0.95 \pm 0.05$ & $0.0075 \pm 0.0007$ & $0.0025 \pm 0.0005$ & $0.01 \pm 0.002$ & $0.3 \pm 0.1$  \\ 
\botrule
\end{tabular}}
\label{tab:cs}
\end{table}

\section{Conclusion}

Let us summarize briefly the results of our work.

It was shown that the disagreement between theoretical predictions and experimental results in $z$-distribution of prompt $J/\psi$ meson production in jets can be removed if one takes into account evolution of the fragmentation function $c\to J/\psi +c$ and double parton scattering contribution. 

It should be mentioned that in our calculations we restrict ourselves to color-singlet mechanism. In the recent work \cite{Bain:2017wvk} it was proposed to explain the same discrepancy taking into account also contributions of color-octet components of the $J/\psi$ meson. 
A more detailed experimental study if the process under consideration (including, probably, polarization measurements \cite{Kang:2017yde}) could help to solve this problem more  accurately.
In addition, our model predicts that charmed meson should be present comoving with $J/\psi$ with $z(D)\approx z(J/\psi)/2$, so it could be interesting to search for this particle.

It should be noted also, that according to papers \cite{Berezhnoy:1998aa, Baranov:2006dh, Artoisenet:2007xi} for transverse momentum under consideration non-fragmentation contributions could also be important. In our future work we plan to study this question in more details.

The authors would like to thank Dr. Lansberg and Dr. Filippova  for fruitful discussions. This work was partially supported by the Russian Foundation of Basic Research grant \#14-02-00096. 


\begin{thebibliography}{10}
	
	\bibitem{Aaij:2017fak}
	LHCb Collaboration, R.~Aaij {\em et~al.}  (2017),
	\href{http://arxiv.org/abs/1701.05116}{{\ttfamily arXiv:1701.05116
			[hep-ex]}}.
	
	\bibitem{Sjostrand:2007gs}
	T.~Sjostrand, S.~Mrenna and P.~Z. Skands, {\em Comput. Phys. Commun.} {\bf
		178}, 852  (2008), \href{http://arxiv.org/abs/0710.3820}{{\ttfamily
			arXiv:0710.3820 [hep-ph]}}.
	
	\bibitem{Braaten:1993jn}
	E.~Braaten, K.-m. Cheung and T.~C. Yuan, {\em Phys. Rev.} {\bf D48},   R5049
	(1993), \href{http://arxiv.org/abs/hep-ph/9305206}{{\ttfamily
			arXiv:hep-ph/9305206 [hep-ph]}}.
	
	\bibitem{Kiselev:1994qp}
	V.~V. Kiselev, A.~K. Likhoded and M.~V. Shevlyagin, {\em Z. Phys.} {\bf C63},
	77  (1994).
	
	\bibitem{Braaten:1993mp}
	E.~Braaten, K.-m. Cheung and T.~C. Yuan, {\em Phys. Rev.} {\bf D48}, 4230
	(1993), \href{http://arxiv.org/abs/hep-ph/9302307}{{\ttfamily
			arXiv:hep-ph/9302307 [hep-ph]}}.
	
	\bibitem{Berezhnoy:1998aa}
	A.~V. Berezhnoy, V.~V. Kiselev, A.~K. Likhoded and A.~I. Onishchenko, {\em
		Phys. Rev.} {\bf D57}, 4385  (1998),
	\href{http://arxiv.org/abs/hep-ph/9710339}{{\ttfamily arXiv:hep-ph/9710339
			[hep-ph]}}.
	
	\bibitem{Novoselov:2010zz}
	A. Novoselov, {\em Phys. Atom. Nucl.} {\bf 73}, 1740  (2010),
	\href{http://arxiv.org/abs/1007.0846}{{\ttfamily arXiv:1007.0846 [hep-ph]}},
	[Yad. Fiz. 73, 1789(2010)].
	
	\bibitem{Corcella:2007tg}
	G.~Corcella and G.~Ferrera, {\em JHEP} {\bf 12},   029  (2007),
	\href{http://arxiv.org/abs/0706.2357}{{\ttfamily arXiv:0706.2357 [hep-ph]}}.
	
	\bibitem{Gribov:1972ri}
	V.~N. Gribov and L.~N. Lipatov, {\em Sov. J. Nucl. Phys.} {\bf 15}, 438
	(1972), [Yad. Fiz.15,781(1972)].
	
	\bibitem{Lipatov:1974qm}
	L.~N. Lipatov, {\em Sov. J. Nucl. Phys.} {\bf 20}, 94  (1975), [Yad.
	Fiz.20,181(1974)].
	
	\bibitem{Dokshitzer:1977sg}
	Y.~L. Dokshitzer, {\em Sov. Phys. JETP} {\bf 46}, 641  (1977), [Zh. Eksp. Teor.
	Fiz.73,1216(1977)].
	
	\bibitem{Altarelli:1977zs}
	G.~Altarelli and G.~Parisi, {\em Nucl. Phys.} {\bf B126}, 298  (1977).
	
	\bibitem{Bain:2017wvk}
	R.~Bain, L.~Dai, A.~Leibovich, Y.~Makris and T.~Mehen  (2017),
	\href{http://arxiv.org/abs/1702.05525}{{\ttfamily arXiv:1702.05525
			[hep-ph]}}.
	
	\bibitem{Kang:2017yde}
	Z.-B. Kang, J.-W. Qiu, F.~Ringer, H.~Xing and H.~Zhang  (2017),
	\href{http://arxiv.org/abs/1702.03287}{{\ttfamily arXiv:1702.03287
			[hep-ph]}}.
	
	\bibitem{Baranov:2006dh}
	S.~P. Baranov, {\em Phys. Rev.} {\bf D73},   074021  (2006).
	
	\bibitem{Artoisenet:2007xi}
	P.~Artoisenet, J.~P. Lansberg and F.~Maltoni, {\em Phys. Lett.} {\bf B653}, 60
	(2007), \href{http://arxiv.org/abs/hep-ph/0703129}{{\ttfamily
			arXiv:hep-ph/0703129 [hep-ph]}}.
	
\end{thebibliography}

\end{document}